\documentclass[usenatbib]{mn2e}
\usepackage{graphicx}
\usepackage{amsmath,amssymb}

\title[Microlensing of the X-ray, UV and optical emission regions of
quasars]{Microlensing of the X-ray, UV and optical emission regions
of quasars: Simulations of the time-scales and amplitude variations
of microlensing events}
\author[P. Jovanovi\'c, A.F. Zakharov,
L.\v C. Popovi\'c,  T. Petrovi\' c]{P. Jovanovi\'c$^1$,  A. F. Zakharov$^{2,3,4}$, L. \v C. Popovi\'c$^1$, T. Petrovi\' c$^1$ \\
$^1$Astronomical Observatory, Volgina 7, 11160 Belgrade, Serbia \\
$^2$Institute of Theoretical and Experimental Physics,
25, B.Cheremushkinskaya, 25, Moscow, 117259, Russia \\
$^3$Bogoliubov Laboratory for Theoretical Physics, JINR, 141980 Dubna,
Russia \\
$^4$Center of Advanced Mathematics and Physics, National University of Science and Technology, Rawalpindi, Pakistan}
\begin{document}

\date{Accepted 2007 Received 2007 ; in original form 2007}

\pagerange{\pageref{firstpage}--\pageref{lastpage}} \pubyear{2007}

\maketitle

\label{firstpage}

\begin{abstract}
We consider the influence of microlensing on different spectral
bands of lensed QSOs. We assumed that the emitting X-ray, UV and
optical regions are different in size, but that the continuum
emission in these spectral bands is originating from an accretion
disc. Estimations of the time scales for microlensing and flux
amplification in different bands are given. We found that the
microlensing duration should be shorter in the X-ray (several
months) than in UV/optical emitting region (several years). This
result indicates that monitoring of the X-ray variations in lensed
QSOs that show a 'flux anomaly' can clarify the source of this anomaly.
 \end{abstract}


\maketitle

\section{Introduction}

Recent observational and theoretical studies suggest that
gravitational microlensing can induce  variability not only in
optical light, but also in the X-ray emission of lensed QSOs
\citep{Chart02a,Chart04,Dai03,Dai04,pj06,Pop01,Pop03a,Pop03b,Pop06a,Pop06b}.
Variability studies of QSOs indicate  that the size of the X-ray
emitting region is significantly smaller ($\sim$ several light
hours), than the optical and UV emitting regions ($\sim$ several
light days).

Gravitational lensing is achromatic (the deflection angle of a light
ray does not depend on its wavelength), but it is clear that if the
geometries of the emitting regions at different wavelengths are
different then chromatic effects could occur. For example, if the
microlens is a binary star or if the microlensed source is extended
\citep{Griest_Hu92,Griest_Hu93,Bog_Cher95a,Bog_Cher95b,Zakh97,Zakh_Sazh98,Pc04}
different amplifications in different spectral bands can be present.
Studies aiming to determine the influence of microlensing on the
spectra of lensed QSOs need to take into account the complex
structure of the QSO central emitting region \citep{Pc04}. Since the
sizes of the emitting regions are wavelength dependent, microlensing
by stars in the lens galaxy may lead to a wavelength dependent
magnification. For example, Blackburne et al. (2006) reported such a
'flux anomaly' in quadruply imaged quasar 1RXS J1131-1231. In
particular, they found discrepancies between the X-ray and optical
flux ratio anomalies. Such anomalies in the different spectral band
flux ratios can be attributed to micro- or milli-lensing in the
massive lensing halo. In the case of milli-lensing one can infer the
nature of substructure in the lensing galaxy, which can be connected
to Cold Dark Matter (CDM) structures (see e.g. Dobler \& Keeton
2006). Besides microlensing, there are several mechanisms which can
produce flux anomalies, such as extinction and intrinsic
variability. These anomalies were discussed in \citet{Pc04} where
the authors gave a method that can aid in distinguishing between
variations produced by microlensing from ones resulting from other
effects. In this paper we discuss consequences of variations due to
gravitational microlensing, in which the different geometries and
dimensions of emitting regions of different spectral bands are
considered.

The influence of microlensing on QSO spectra emitted from their
accretion discs in the range from the X-ray to the optical
spectral band is analyzed. Moreover, assuming different sizes
of the emitting regions, we investigate the microlensing time
scales for those regions, as well as a time dependent response
in amplification of different spectral bands due to
microlensing. Also, we give the estimates of microlensing time
scales for a sample of lensed QSOs.

In Section 2, we describe our model of the quasar emitting regions
and a model of the micro-lens. In Section 3 we discuss the time
scales of microlensing and in Section 4 we present our results.
Finally, in Section 5, we draw our conclusions.

\section{A model of QSO emitting regions and microlens}

\subsection{A model of the QSO emitting regions}

In our models we adopt a disc geometry for the emitting regions of
Active Galactic Nuclei (AGN)  since the most widely accepted
paradigm for AGN includes a supermassive black hole fed by an
accretion disc. \citet{Fabian89} calculated spectral line profiles
for radiation emitted from the inner parts of accretion discs and
later on such features of  Fe $K\alpha$ lines were discovered by
\citet{Tanaka95} in Japanese ASCA satellite data for Seyfert galaxy
MGC-6-30-15. Moreover, the assumption of a disc geometry for the
distribution of the emitters in the central part is supported by the
spectral shape of the Fe K$\alpha$ line in AGN (e.g.
\citet{Nandra_1997}, see also results of simulations
\citep{Zak_Rep99,Zak_Rep02,ZKLR02,Zak_Rep03a,Zak_Rep03b,ZR_Nuovo_Cim03,Zak_Rep03c,ZR_5SCSLSA,ZR_NA_05,Zak_SPIG04,ZR_NANP_05}).
On the other hand, very often a bump in the UV band is present in
the spectra of AGN, that indicates that the UV and optical continuum
originates in an accretion disk.

We should note here that probably most of the X-ray emission in the
1--10 keV energy range originates from inverse Compton scattering of
photons from the disc by electrons in a tenuous hot corona. Proposed
geometries of the hot corona of AGN include  a spherical corona
sandwiching the disc and a patchy  corona made of a few compact
regions covering a small fraction of the disc (see e.g.
\citet{mz07}). On the other hand, it is known that part of the
accretion disc that emits in the 1--10~keV rest-frame band (e.g. the
region that emits the continuum Compton reflection component and the
fluorescent emission lines) is very compact and may contribute to
X-ray variability in this energy range. In order to study the
microlensing time scales, one should consider the dimensions of the
X-ray emitting region which are very important for the integral flux
variations due to microlensing. The geometry of the emitting regions
adopted in microlensing models will affect the simulated spectra of
the continuum and spectral line emission \citep{Pc04,pj06}. In this
paper we assume that the AGN emission from the optical to the X-ray
band originates from different parts of the accretion disc.

Also we assume that we have a stratification in the disc, where the
innermost part emits the X-ray radiation and outer parts the UV and
optical continuum emission. To study the effects of microlensing on
a compact accretion disc we use the ray-tracing method, considering
only those photon trajectories that reach the sky plane at a given
observer angle (see, e.g. Popovi\'c et al. 2003a,b and references
therein). The amplified brightness with amplification $A(X,Y)$ for
the continuum is given by

\begin{equation}
I_{C} (X,Y;E_{obs}) = { {I_{P}}}  (E_{obs},T(X,Y)) \cdot A(X,Y),
\end{equation}
where $T(X,Y)$ is the temperature, $X$ and $Y$ are the impact
parameters which describe the apparent position of each point of the
accretion disc image on the celestial sphere as seen by an observer
at infinity, $E_{\rm obs}$ is the observed energy, $I_P$ is an
emissivity function.

In the standard Shakura-Sunyaev disc model \citeyearpar{Shakura73},
accretion occurs via an optically thick and geometrically thin disc.
The effective optical depth in the disc is very high and photons are
close to the thermal equilibrium with electrons. The surface
temperature is a function of disc parameters and results in the
multicolor black body spectrum. This component is thought to explain
the 'blue bump' in AGN and the soft X-ray emission in galactic black
holes. Although the standard Shakura-Sunyaev disc model does not
predict the power-law X-ray emission observed in all sub-Eddington
accreting black holes, the power law for the X-ray emissivity in AGN
is usually adopted (see e.g. \cite{Nandra_1999}). But one can not
exclude other approximations for emissivity laws, such as black-body
or modified black-body emissivity laws. Moreover, we will assume
that the emissivity law is the same through the whole disc.
Therefore we used the black-body radiation law. The disc emissivity
is given as (e.g. \cite{Jarosz_1992}):

$$I_P(X,Y;E)=B[E,T_s(X,Y)],$$
where
\begin{equation}
 B\left( {E,T_s(X,Y)} \right) = {\frac{{2E
^{3}}}{{h^2c^{2}}}}{\frac{{1}}{{e^{{{E
}
\mathord{\left/ {\vphantom {{h\nu}  {kT}}} \right.
\kern-\nulldelimiterspace} {kT_s(X,Y)}}} - 1}}},
\end{equation}
where $c$ is the speed of light, $h$ is the Planck constant,
$k$ is the Boltzmann constant and $T_s(X, Y)$ is the  surface
temperature. Concerning the standard accretion disc
\citep{Shakura73}, here we assumed that (see \citet{Pop06a}):
\begin{equation}
T_s(X, Y) \sim r^{-3/2}(X,Y)(1-r^{-1/2}(X,Y))^{4/5} \,{\rm K},
\end{equation}
taking that an effective temperature in the innermost part (where X-ray is
originated) is in an interval from 10$^7$ K to
10$^8$ K.

\begin{figure*}
\centering
\includegraphics[width=0.49\textwidth]{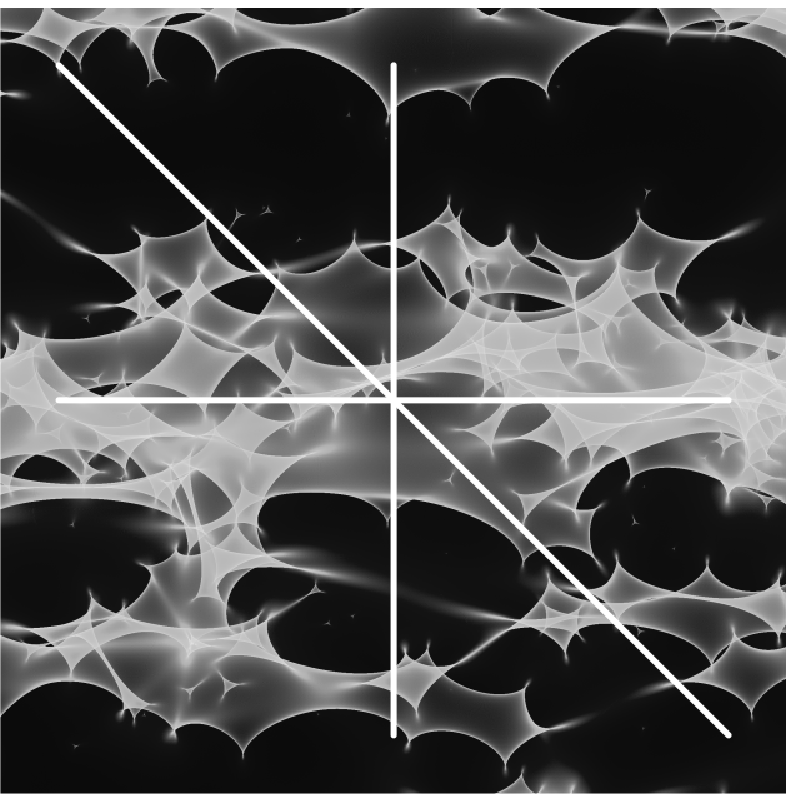}\hfill
\includegraphics[width=0.49\textwidth]{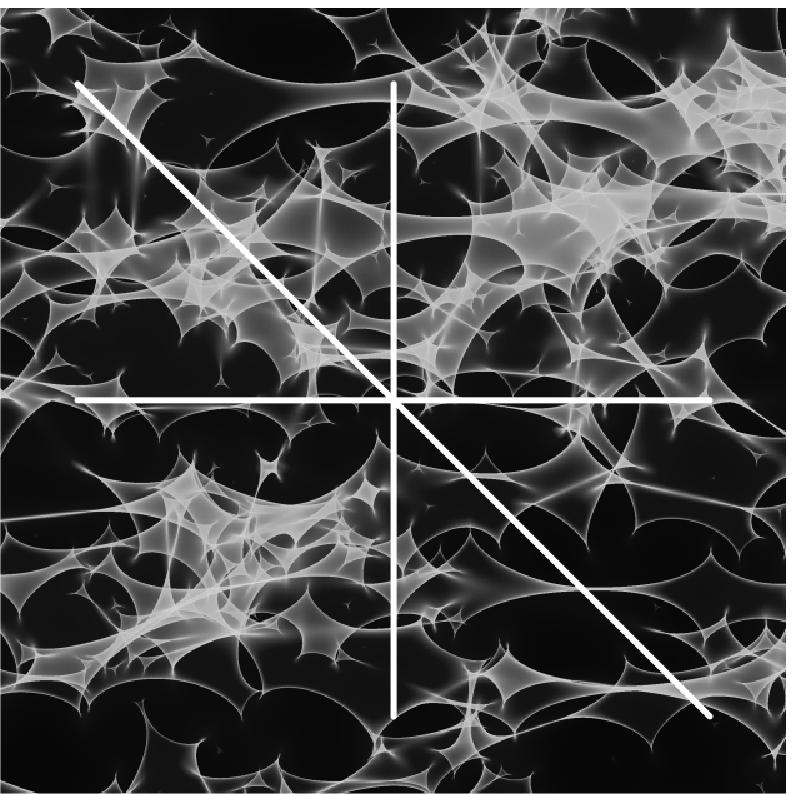}
\caption{Magnification map of the QSO 2237+0305A image (left) and of
a "typical" lens system (right). The white solid lines represent
three analyzed paths of an accretion disc center: horizontal
($y=0$), diagonal ($y = -x$) and vertical ($x=0$).}
\end{figure*}

\subsection{A model for microlensing}

To explain the observed microlensing events in quasars, one can use
different microlensing models. The simplest approximation is a
point-like microlens, where microlensing is caused by some compact
isolated object (e.g. by a star). Such a microlens is characterized
by its Einstein Ring Radius in the lens plane:
$ERR=\sqrt{\dfrac{4Gm}{c^2}\dfrac{D_lD_{ls}}{D_s}}$ or by the
corresponding projection in the source plane:
$R_E=\dfrac{D_s}{D_l}ERR=\sqrt{\dfrac{4Gm}{c^2}\dfrac{D_sD_{ls}}{D_l}}$,
where $G$ is the gravitational constant, $c$ is the speed of light,
$m$ is the microlens mass and $D_l$, $D_s$ and $D_{ls}$ are the
cosmological angular distances between observer-lens,
observer-source and lens-source, respectively. In most cases we can
not simply consider that microlensing is caused by an isolated
compact object but we must take into account that the
micro-deflector is located in an extended object (typically, the
lens galaxy). Therefore, when the size of the Einstein ring radius
projection $R_E$ of the microlens is larger than the size of the
accretion disc and when a number of microlenses form a caustic net,
one can describe the microlensing in terms of the crossing of the
disc by a straight fold caustic (Schneider et al. 1992). The
amplification at a point of an extended source (accretion disc)
close to the caustic is given by \citet{Chang84} (a more general
expression for a magnification near a cusp type singularity is given
by \citet{Schneider92,Zakharov95}):
\begin{equation}
A(X,Y)=A_0+K\sqrt{\frac{r_{\rm caustic}}{\kappa(\xi-\xi_c)}}\cdot
H(\kappa(\xi-\xi_c)),
\label{eq04}
\end{equation}
where $A_0$ is the amplification outside  the caustic and
$K=A_0\beta$ is the caustic amplification factor, where $\beta$ is
constant of order of unity (e.g. Witt et al. 1993). The "caustic
size" $r_{\rm caustic}$ is the distance in a direction
perpendicular to the caustic for which the caustic amplification is
1, and therefore this parameter defines a typical linear scale for
the caustic in the direction perpendicular to the caustic. $\xi$ is
the distance perpendicular to the caustic in gravitational radii
units and $\xi_c$ is the minimum distance from the disc center to
the caustic. Thus,
\begin{equation}
\xi_c={\sqrt{X_c^2+Y_c^2}},
\end{equation}

\begin{equation}
{\rm tg}\alpha=\frac{Y_c}{X_c},
\end{equation}
 and
\begin{equation}
\xi=\xi_c+\frac{(X-X_c){\rm tg}\phi+Y_c-Y}{\sqrt{{\rm tg}^2\phi+1}},
\end{equation}
 where
$\phi=\alpha+{\pi/2}$. $ H(\kappa(\xi-\xi_c))$ is the Heaviside
function, $H(\kappa(\xi-\xi_c))=1$, for $\kappa(\xi-\xi_c)>0$,
otherwise it is 0. $\kappa$ is $\pm 1$, it depends on the direction
of caustic motion; if the direction of the caustic motion is from
approaching side of the disc $\kappa=-1$, otherwise it is +1. Also,
in the special case of caustic crossing perpendicular to the
rotating axis $\kappa=+1$ for direction of caustic motion from $-Y$
to $+Y$, otherwise it is $-1$. A microlensing event where a caustic
crosses over an emission region can be described in the following
way: before the caustic reaches the emission region, the
amplification is equal to $A_0$ because the Heavisied function of
equation (4) is zero. Just as the caustic begins to cross the
emitting region the amplification rises rapidly and then decays
gradually towards $A_0$ as the source moves away from the
caustic-fold.

Moreover, for the specific event one can model the caustic shape to
obtain different parameters (see e.g. Abajas et al. 2005, Kochanek
2004 for the case of Q2237+0305). In order to apply an appropriate
microlens model, additionally we will consider a standard
microlensing magnification pattern for the Q2237+0305A image (Fig. 1
left). For generating this map we used the ray-shooting method
\citep{Kay86,sch1,sch2,WP90,WSP90}. In this method the input
parameters are the average surface mass density $\sigma$, shear
$\gamma$ and width of the microlensing magnification map expressed
in units of the Einstein ring radius (defined for one solar mass in
the lens plane).

First, we generate a random star field in the lens plane with use of
the parameter $\sigma$. After that, we solve the Poisson equation
$\nabla^2\psi=2\sigma$ in the lens plane numerically, so we can
determine the lens potential $\psi$ in every point of the grid in
the lens plane. To solve the Poisson equation numerically one has to
write its finite difference form:
\begin{equation}
\psi_{i+1,j}+\psi_{i-1,j}+\psi_{i,j+1}+\psi_{i,j-1}-4\psi_{i,j}=2\sigma_{i,j}.
\label{psi}
\end{equation}
Here we used the standard 5-point formula for the two-dimensional
Laplacian.  Next step is inversion of the equation (\ref{psi}) using
Fourier transform. After some transformations we obtain:
\begin{equation}
\hat{\psi}=\frac{\hat{\sigma}_{mn}}{2(\cos{\frac{m\pi}{N_{1}}}+\cos{\frac{n\pi}{N_{2}}}-2)},
\label{hatpsi}
\end{equation}
where $N_1$ and $N_2$ are dimensions of the grid in the lens plane.
Now, using the finite difference technique, we can compute the
deflection angle $ \vec{\alpha}=\nabla\psi $ in each point of the
grid in the lens plane. After computing deflection angle, we can map
the regular grid of points in the lens plane, via lens equation,
onto the source plane. These light rays are then collected in pixels
in the source plane, and the number of rays in one pixel is
proportional to the magnification due to microlensing at this point
in the source plane.

Typically, for calculations of microlensing in
gravitationally lensed systems one can consider cases where
dimensionless surface mass density $\sigma$ is some fraction of
1 e.g. 0.2, 0.4, 0.6, 0.8 \citep{Treyer03}, cases without
external shear ($\gamma=0$) and cases with $\gamma=\sigma$ for
isothermal sphere model for the lensing galaxy. In this article
we assume the values of these parameters within the range
generally adopted by other authors, in particular by
\citet{Treyer03}.

Microlensing magnification pattern for the Q2237+0305A image (Fig. 1
left) with 16~$R_E$ on a side (where $R_E\approx 5867\ R_g$) is
calculated using the following parameters: $\sigma=0.36$ and
$\gamma=0.4$ (see \cite{Pop06a}, Fig. 2), the mass of microlens is
taken to be 0.3$M_\odot$ and we assume a flat cosmological model
with $\Omega =0.3$  and $H_{0}= 75\ \rm km\ s^{-1} Mpc^{-1}$.

We also calculated microlensing magnification pattern for a
"typical" lens system (Fig. 1 right), where the redshifts of
microlens and source are: $z_l=0.5$ and $z_s=2$. In this case, the
microlens parameters are taken arbitrary: $\sigma=0.45$ and
$\gamma=0.3$ and the size of obtained microlensing pattern is also
$16\ R_E \times 16\ R_E$, where $R_E\approx 3107\ R_g$.

\section{Typical time scales for microlensing}

Typical scales for microlensing are discussed not only in books on
gravitational lensing \citep{Schneider92a,Zakh97,Petters01}, but in
recent papers also (see, for example, \cite{Treyer03}). In this
paper we discuss microlenses located in gravitational macrolenses
(stars in lensing galaxies), since optical depth for microlensing is
then the highest \citep{Wyithe02,Wyithe02b,Zakharov04,Zakharov05} in
comparison with other possible locations of gravitational
microlenses, as for example stars situated in galactic clusters and
extragalactic dark halos \citep{Tadros98,Totani03,Inoue03}.

Assuming the concordance cosmological model with $\Omega_{\rm tot}=1$,
$\Omega_{\rm matter}=0.3$ and $\Omega_{\Lambda}=0.7$ we
recall that typical length scale for microlensing is \citep{Treyer03}:
\begin{equation}
 R_E = \sqrt{2 r_s \cdot  \frac{D_s D_{ls}}{D_{l}}}
 \approx 3.2 \cdot 10^{16} \sqrt{\frac{m}{M_\odot}} h_{75}^{-0.5} \mathrm{~cm},
 \label{eq_suppl1}
\end{equation}
where "typical" microlens and sources redshifts are assumed to be
$z_l=0.5, z_s=2.$ (similar to \cite{Treyer03}),
$r_s=\dfrac{2Gm}{c^2}$ is the Schwarzschild radius corresponding to
microlens mass $m$, $h_{75}=H_0/((75 {\rm~km/sec})/{\rm Mpc})$ is
dimensionless Hubble constant.

The corresponding angular scale is \citep{Treyer03}
\begin{equation}
\theta_0=\frac{R_E}{D_s}
 \approx 2.2 \cdot 10^{-6} \sqrt{\frac{m}{M_\odot}} h_{75}^{-0.5}{\rm ~arcsec},
 \label{eq_suppl2}
\end{equation}

Using the length scale (\ref{eq_suppl1}) and velocity scale
(say $v_\bot \sim$~600~km/sec as \cite{Treyer03} did), one
could calculate the standard time scale corresponding to the
scale to cross the projected Einstein radius
\begin{equation}
t_E=(1+z_l)\frac{R_E}{v_\bot}
 \approx 25 \sqrt{\frac{m}{M_\odot}}v_{600}^{-1} h_{75}^{-0.5}{\rm ~years},
 \label{eq_suppl3}
\end{equation}
where a relative transverse velocity $v_{600}=v_\bot/(600$~km/sec).
The time scale $t_E$, corresponding to a point-mass lens and to a
small source (compared to the projected Einstein radius of the
lens), could be used if microlenses are distributed freely at
cosmological distances and if each Einstein angle is located far
enough from another one. However, the estimation (\ref{eq_suppl3})
gives long and most likely overestimated time scales especially for
gravitationally lensed systems. Thus we must apply another microlens
model to estimate time scales.

For a simple caustic model, such as one that considers a straight
fold caustic\footnote{We use the following approximation for the
extra magnification near the caustic: $\mu=\sqrt{\dfrac{r_{\rm
caustic}}{\xi-\xi_c}}$, $\xi>\xi_c$ where $\xi$ is the perpendicular
direction to the caustic fold (it is obtained from Eq. (\ref{eq04})
assuming that factor $K$ is about unity).}, there are two time
scales depending either on the "caustic size" ($r_{\rm caustic}$) or
the source radius ($R_{\rm source}$). In the case when source radius
is larger or at least close to the "caustic size" ($R_{\rm source}
\gtrsim r_{\rm caustic}$), the relevant time scale is the "crossing
time" \citep{Treyer03}:
\begin{eqnarray}
\label{eq_suppl4}
t_{\rm cross} & = & (1+z_l)\frac{R_{\rm source}}{v_\bot (D_s/D_l)} \nonumber \\
& \approx & 0.69\ R_{15}\ v_{600}^{-1} \left(\frac{D_s}{D_l}\right)^{-1} h_{75}^{-0.5}{\rm ~years} \\
& \approx & 251\ R_{15}\ v_{600}^{-1}\ h_{75}^{-0.5}{\rm ~days},\nonumber
\end{eqnarray}
where $D_l$ and $D_s$ correspond to $z_l=0.5$ and $z_s=2$,
respectively and $R_{15}=R_{\rm source}/10^{15}$~cm. As a matter of
fact, the velocity perpendicular to the straight fold caustic
characterizes the time scale and it is equal to $v_\bot \sin \beta$
where $\beta$ is the angle between the caustic and the velocity
$v_\bot$ in the lens plane, but in our rough estimates we can omit
factor $\sin \beta$ which is about unity. However, if the source
radius $R_{\rm source}$ is much smaller than the "caustic size"
$r_{\rm caustic}$ ($R_{\rm source} \ll r_{\rm caustic}$), one could
use the "caustic time", i.e. the time when the source is located in
the area near the caustic:
\begin{eqnarray}
\label{eq_suppl5}
t_{\rm caustic} & = & (1+z_l)\frac{r_{\rm caustic}}{v_\bot (D_s/D_l)} \nonumber \\
& \approx & 0.69\ r_{15}\ v_{600}^{-1} \left(\frac{D_s}{D_l}\right)^{-1} h_{75}^{-0.5}{\rm ~years} \\
& \approx & 251\ r_{15}\ v_{600}^{-1}\ h_{75}^{-0.5}{\rm ~days}, \nonumber
\end{eqnarray}
where $r_{15}=r_{\rm caustic}/10^{15}$~cm.

Therefore, $t_{\rm cross}$ could be used as a lower limit for
typical time scales in the case of a simple caustic microlens model.
From equations (\ref{eq_suppl4}) and (\ref{eq_suppl5}) it is clear
that one cannot unambiguously infer the source size $R_{\rm source}$
from variability measurements alone, without making some further
assumptions. In general, however, we expect that $t_{\rm cross}$
corresponds to smaller amplitude variations than $t_{\rm caustic}$,
since in the first case only a fraction of a source is significantly
amplified by a caustic (due to assumption that $R_{\rm source}
\gtrsim r_{\rm caustic}$), while in the second case it is likely
that the entire source could be strongly affected by caustic
amplification (due to assumption that $R_{\rm source} \ll r_{\rm
caustic}$).

\begin{figure}
\centering
\includegraphics[width=8cm]{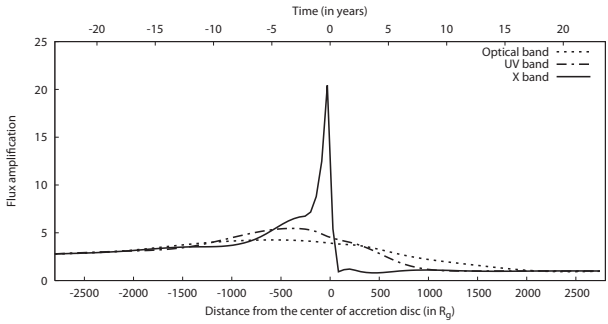}
\caption{The variations of normalized total continuum flux in
optical (3500--7000 \AA), UV (1000--3500 \AA) and X (1.24--12.4 \AA
\ i.e. 1--10 KeV) band due to microlensing by a caustic crossing
along $y=-x$ direction in the case of Schwarzschild metric. Time
scale corresponds to "typical" redshifts of microlens and source:
$z_l=0.5$ and $z_s=2$. The parameters of the caustic are: $A_0$=1,
$\beta$=1, $\kappa=+1$ and its "size" is 9000 $R_g$. Negative
distances and times correspond to the approaching side, and positive
to the receding side of accretion disc. In this case, due to
$\kappa=+1$, caustic motion is from the receding towards the
approaching side (i.e. from the right to the left). The source mass
is $10^8\ M_\odot$. The radii of optical emitting region are:
$R_{in}= 100\ R_g$, $R_{out}=2000\ R_g$, for UV emitting region:
$R_{in}= 100\ R_g$, $R_{out}=1000\ R_g$ and for X emitting region:
$R_{in}= R_{ms}$, $R_{out}=80\ R_g$.}
\end{figure}

\begin{figure}
\centering
\includegraphics[width=8cm]{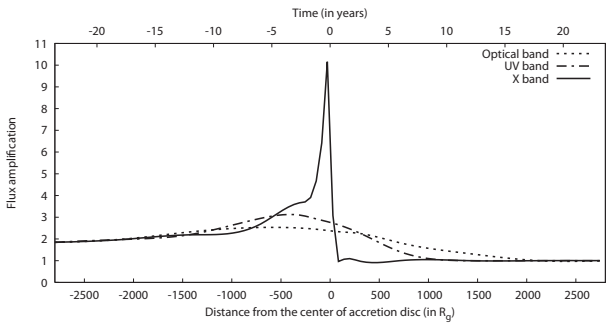}
\caption{The same as in Fig. 2. but for $r_{\rm caustic}=2000\ R_g$.}
\end{figure}

\begin{figure}
\centering
\includegraphics[width=8cm]{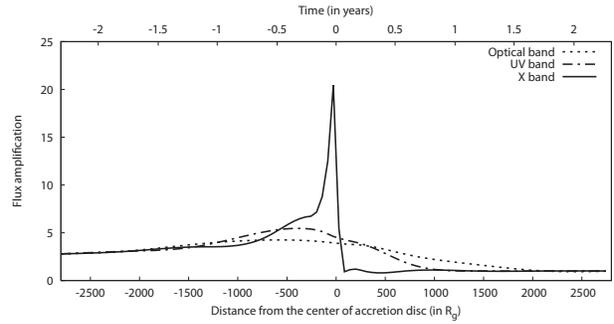}
\caption{The same as in Fig. 2. in the case when time scale
corresponds to $z_l=0.04$, $z_s=1.69$ (i.e. to Q2237+0305 lens
system).}
\end{figure}

In this paper, we estimated the microlensing time scales for the
X-ray, UV and optical emitting regions of the accretion disc using
the following three methods:
\begin{enumerate}
\item by converting the distance scales of microlensing events
    to the corresponding time scales according to the formula
    (13) in which $R_{\rm source}$ is replaced by the distance
    from the center of accretion disc. Caustic rise times
    ($t_{HME}$) are then derived from the simulated variations
    of the normalized total flux in the X-ray, UV and optical
    spectral bands (see Figs. 2-4) by measuring the time from
    the beginning to the peak of the magnification event
\item by calculating the caustic times ($t_{\rm caustic}$)
    according to equation (14)
\item using light curves (see Figs. 5 and 6) produced when the
    source crosses over a magnification pattern. Rise times of
    high magnification events ($t_{HME}$) are then measured as
    the time intervals between the beginning and the maximum of
    the corresponding microlensing events (for more details, see
    the next section).
\end{enumerate}

\section{Results and discussion}

In order to explore different cases of microlensing and evaluate
time scales for different spectral bands, first we numerically
simulate the crossing of a straight-fold caustic with parameters
$A_0=1,\ \beta=1$, $\kappa=+1$ and $r_{\rm caustic}=9000$ $R_g$ over
an accretion disc with an inclination angle 35$^\circ$, that is
stratified into three parts:

(i) The innermost part that emits X-ray continuum (1.24 \AA\ -- 12.4
\AA \ or 1--10 keV). The inner radius is taken to be
$R_{inn}=R_{ms}$ (where $R_{ms}$ is the radius of the marginally
stable orbit: $R_{ms}=6\ R_g$ in the Schwarzschild metric) and outer
radius is $R_{out}=80$ $R_g$ (where $R_g=GM/c^2$ is the
gravitational radius for a black hole with mass $M$).

(ii) An UV emitting part of the disc (contribute to the emission
from 1000 \AA\ -- 3500\AA ), with $R_{inn}=100$ $R_g$ and
$R_{out}=1000$ $R_g$.

(iii) An outer optical part of the disc with $R_{inn}=100$ $R_g$ and
$R_{out}=2000$ $R_g$ that emits in the wavelength band from 3500
\AA\ until 7000 \AA.

\begin{table*}
\centering \caption{The estimated time scales (in years) for
microlensing of the X-ray, UV and optical emission region for lensed
QSOs observed by Chandra X-ray Observatory \citep{Dai04}. The
calculated caustic times $t_{\rm caustic}$ are obtained according to
formula (14) for the following values of the cosmological constants:
$H_0=75\rm \ km\ s^{-1}Mpc^{-1}$ and $\Omega_0=0.3$. The caustic
rise times $t_{HME}$ are derived from caustic crossing simulations
(see Figs 2 -- 4). The black hole mass is assumed to be 10$^8\rm
M_\odot$.}
\begin{tabular}{|c|c|c|c|c|c|c|c|c|}
\hline
 Object & $z_s$ & $z_l$ & \multicolumn{2}{c|}{X-ray} & \multicolumn{2}{c|}{UV}    & \multicolumn{2}{c|}{optical} \\
\cline{4-9}
        &       &       &  $t_{\rm caustic}$  & $t_{HME}$  & $t_{\rm caustic}$  & $t_{HME}$ & $t_{\rm caustic}$ & $t_{HME}$  \\
\hline
\hline
    HS 0818+1227   &  3.115 & 0.39 & 0.572 & 0.660 & 7.147  &  7.070 & 14.293 & 15.160 \\
  RXJ 0911.4+0551  &  2.800 & 0.77 & 0.976 & 1.120 & 12.200 & 12.080 & 24.399 & 25.880 \\
   LBQS 1009-0252  &  2.740 & 0.88 & 1.077 & 1.240 & 13.468 & 13.330 & 26.935 & 28.570 \\
     HE 1104-1805  &  2.303 & 0.73 & 0.918 & 1.050 & 11.479 & 11.370 & 22.957 & 24.350 \\
      PG 1115+080  &  1.720 & 0.31 & 0.451 & 0.520 & 5.634  &  5.570 & 11.269 & 11.950 \\
     HE 2149-2745  &  2.033 & 0.50 & 0.675 & 0.780 & 8.436  &  8.350 & 16.871 & 17.890 \\
      Q 2237+0305  &  1.695 & 0.04 & 0.066 & 0.080 & 0.828  &  0.820 & 1.655  &  1.760 \\
\hline
\end{tabular}
\end{table*}

\begin{table*}
\centering \caption{Average rise times ($<t_{HME}>$) of high
magnification events, their average number ($<N_{\rm caustic}>_0$)
per unit length ($R_E$) and their average number ($<N_{\rm
caustic}>_y$) per year in the light curves of Q2237+0305A (Fig. 5)
and "typical" lens system (Fig. 6).}
\begin{tabular}{|c|c|c|c|c|c|c|c|c|}
\hline
& & \multicolumn{3}{|c|}{Q2237+0305} & \multicolumn{3}{c|}{"Typical" lens} \\
\hline
Disc path & Sp. band & $<t_{HME}>$ & $<N_{\rm caustic}>_0$ & $<N_{\rm caustic}>_y$ & $<t_{HME}>$ & $<N_{\rm caustic}>_0$ & $<N_{\rm caustic}>_y$ \\
\hline
\hline
      & X        & 0.37 & 0.95 & 0.20 & 3.94 & 1.63 & 0.06 \\
$y=0$ & UV       & 1.29 & 0.59 & 0.12 &10.96 & 0.78 & 0.03 \\
      & Optical  & 2.96 & 0.51 & 0.10 &15.81 & 0.62 & 0.02 \\
\hline
      & X        & 0.57 & 0.52 & 0.11 & 2.77 & 1.65 & 0.06 \\
$y=-x$ & UV       & 1.92 & 0.36 & 0.07 & 6.62 & 0.49 & 0.02 \\
      & Optical  & 3.98 & 0.26 & 0.05 &17.04 & 0.38 & 0.01 \\
\hline
      & X        & 0.68 & 1.61 & 0.33 & 4.52 & 1.40 & 0.05 \\
$x=0$ & UV       & 1.38 & 0.88 & 0.18 &13.25 & 0.54 & 0.02 \\
      & Optical  & 2.96 & 0.44 & 0.09 &31.62 & 0.31 & 0.01 \\
\hline
\end{tabular}
\end{table*}

Having in mind that the aims of this investigation are to study the
microlensing time scales for different emitting regions and time
dependent response of amplification in different spectral bands, we
considered microlensing magnification patterns only for image A of
Q2237+0305 and for a "typical" lens system. Our intention was not to
create a complete microlensing model for a specific lens system, and
therefore we did not analyze the differences between images (as for
instance the time delay between them). The variations in the total
flux in the different spectral bands are given in Figs. 2--6. In
Figs. 2 and 3 the simulations for a typical lens system with
$z_l=0.5$ and $z_s=2$ and for two different "caustic sizes" are
given. As one can see from Figs. 2 and 3, the microlensing time
scales are different for different regions, and the durations of
variations in the X-ray are on order of several months to a few
years, but in the UV/optical emission regions they are on order of
several years. Also, as one can see in Figs. 2 and 3, the time
scales do not depend on "caustic size" which, on the other hand,
affects only the maximal amplifications in all three spectral bands.
The results corresponding to the lens system of QSO 2237+0305
($z_l=0.04$, $z_s=1.69$) are given in Figs. 4 and 5. We considered
the straight-fold caustic (Fig. 4) and a microlensing pattern for
the image A of QSO 2237+0305 (Fig. 5). As one can see from Figs. 4
and 5, a higher amplification in the X-ray continuum than in the
UV/optical is expected. In this case, the corresponding time scales
are much shorter and they are from a few days up to a few months for
X-ray and a few years for UV/optical spectral bands. The similar
conclusion arises when we compare the latter results with those for
a "typical" lens system in the case of microlensing pattern (Fig.
6).

\begin{figure*}
\centering
\includegraphics[width=0.45\textwidth]{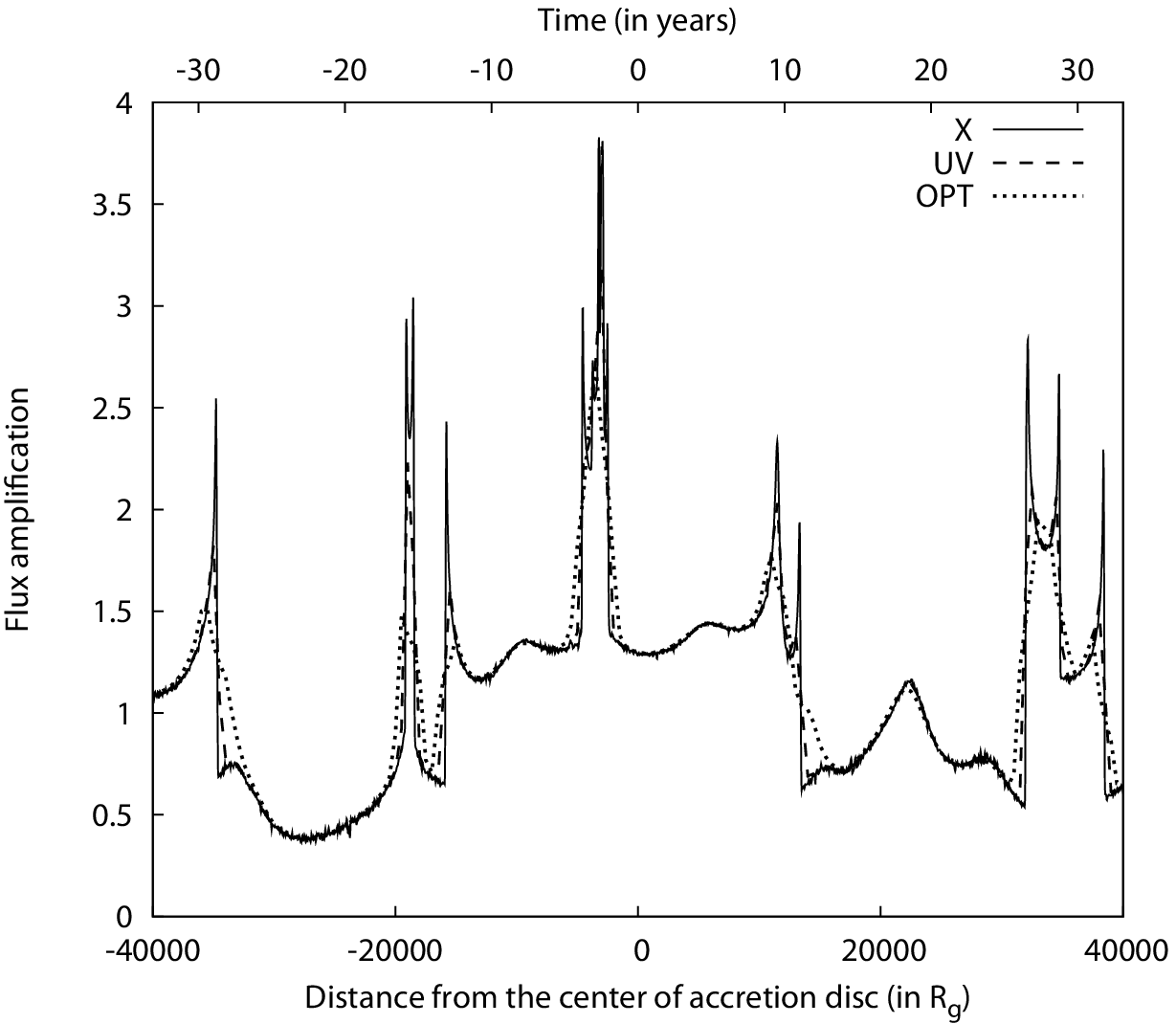}
\includegraphics[width=0.45\textwidth]{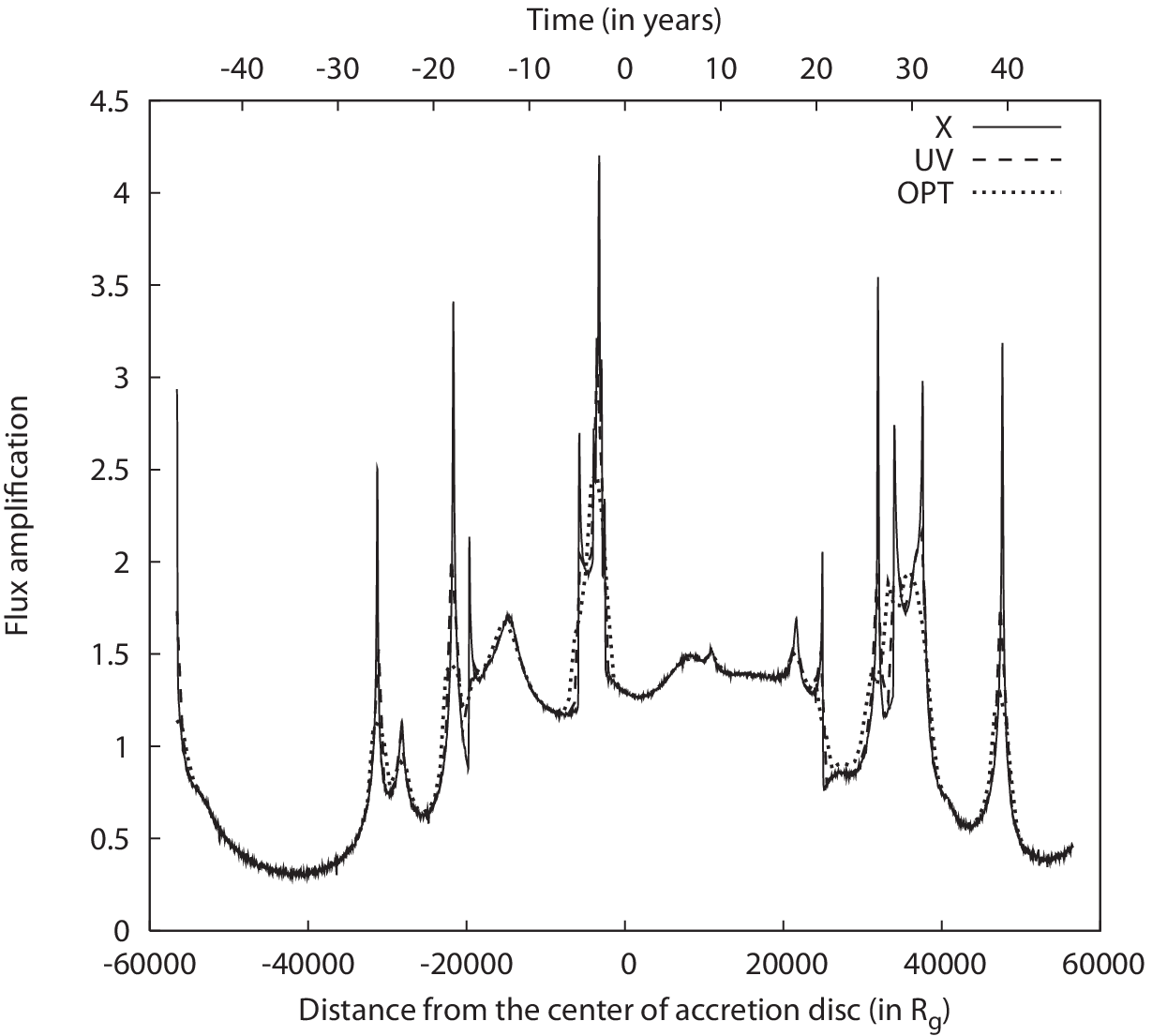} \\
\includegraphics[width=0.45\textwidth]{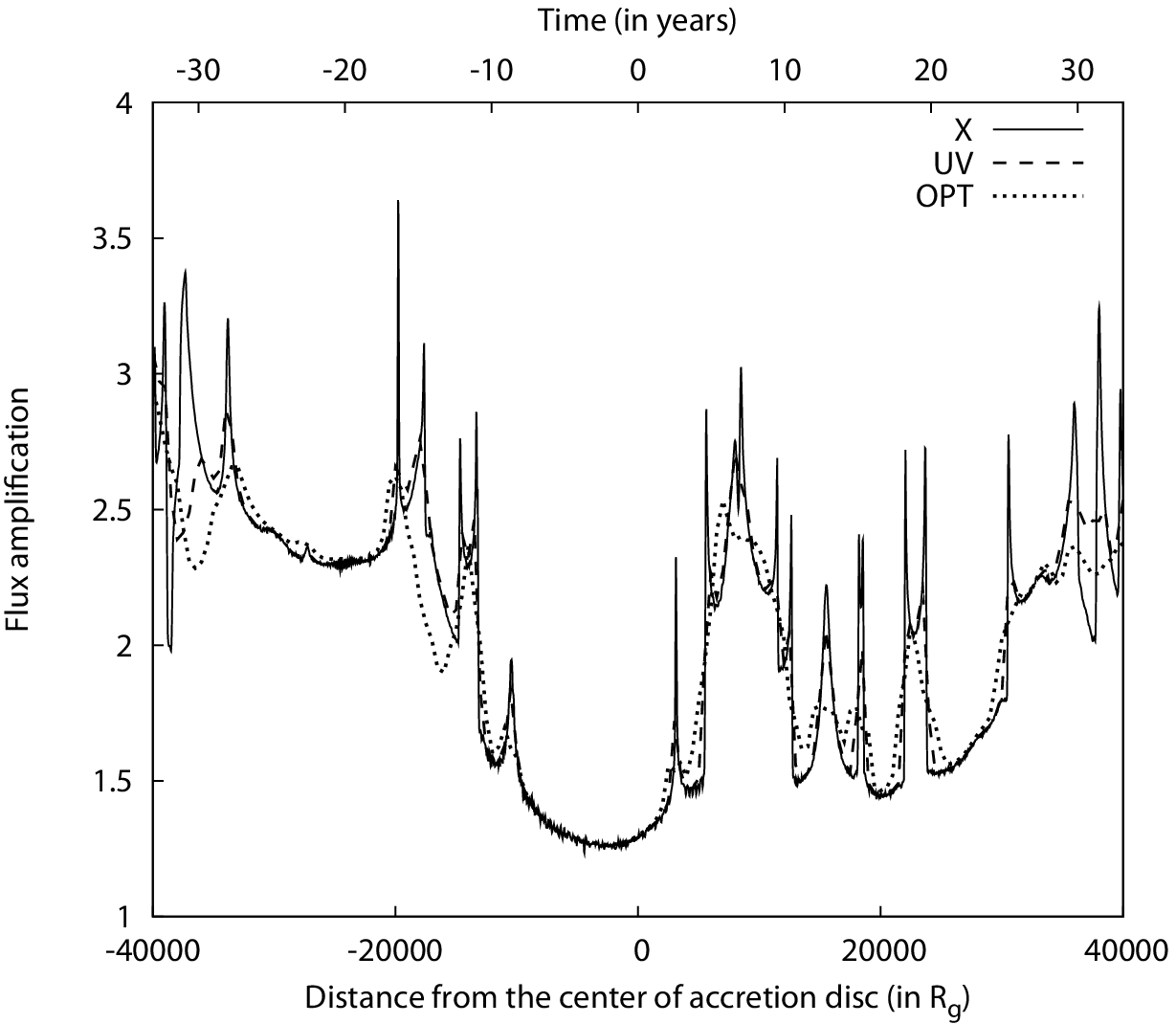}
\caption{Variations in the X-ray (solid), UV (dashed) and optical
(dotted) spectral bands corresponding to the horizontal (top left),
diagonal (top right) and vertical (bottom) path in the magnification
map of the QSO 2237+0305A (Fig. 1 left).}
\end{figure*}

%\clearpage

\begin{figure*}
\centering
\includegraphics[width=0.45\textwidth]{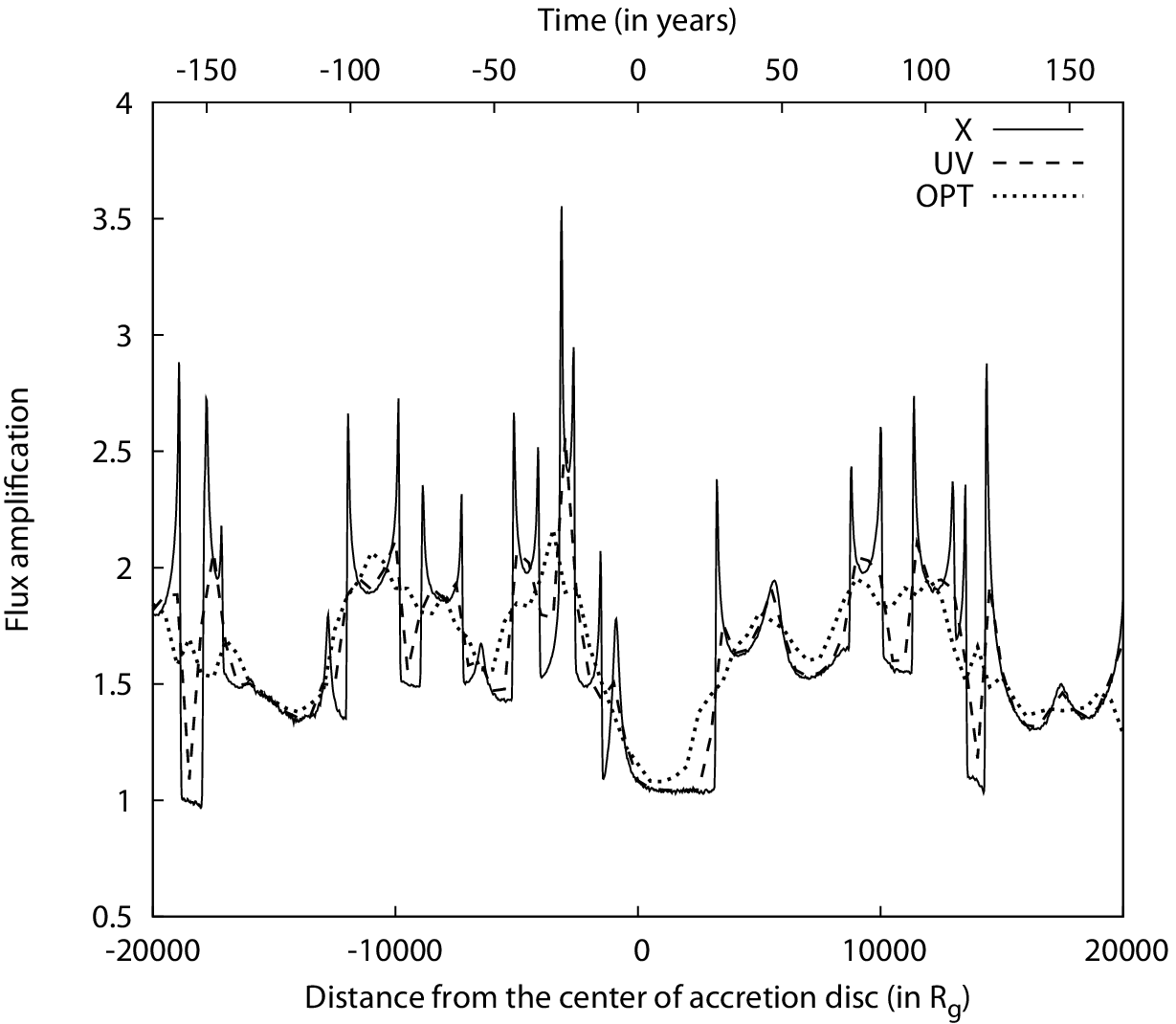}
\includegraphics[width=0.45\textwidth]{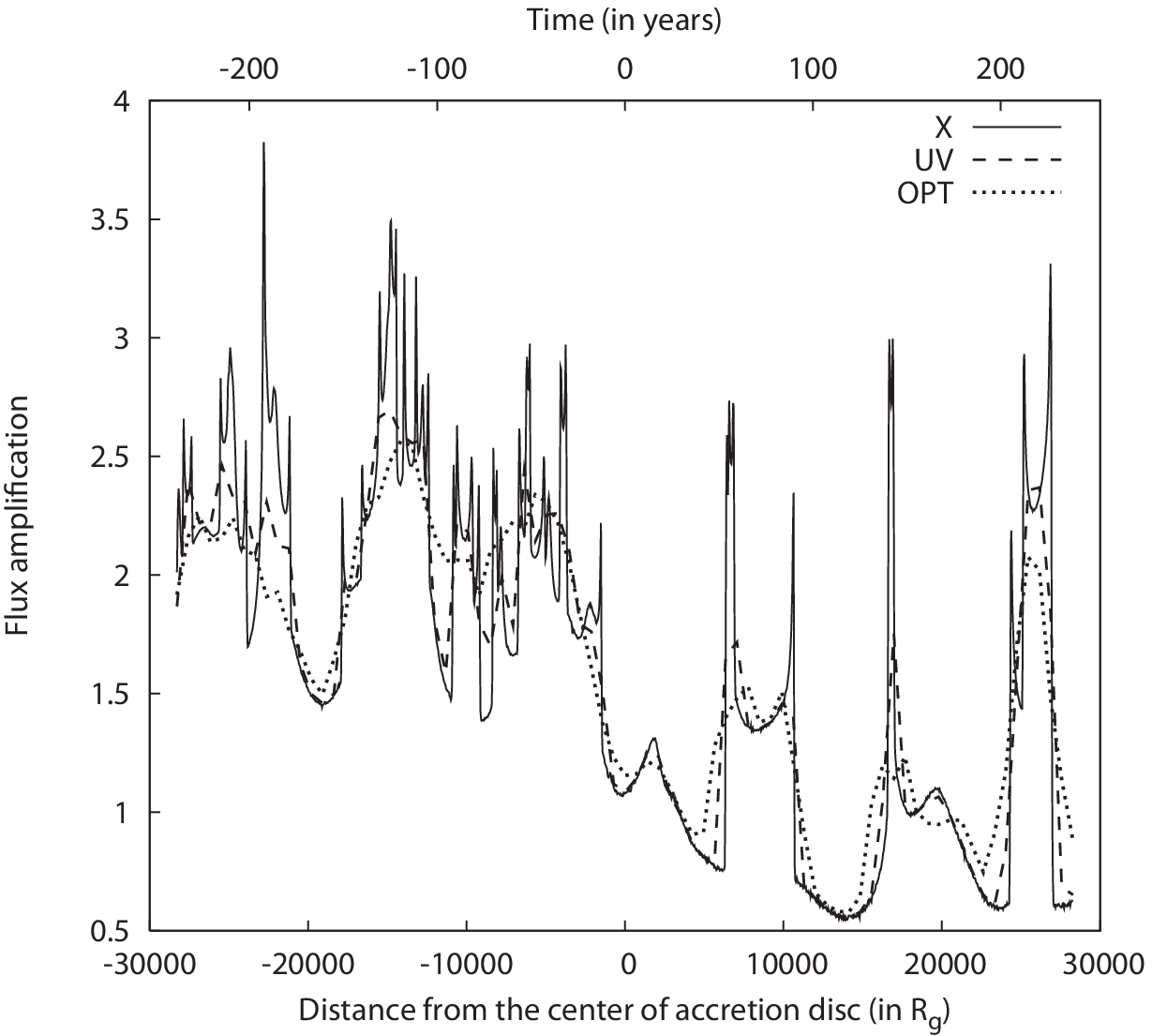} \\
\includegraphics[width=0.45\textwidth]{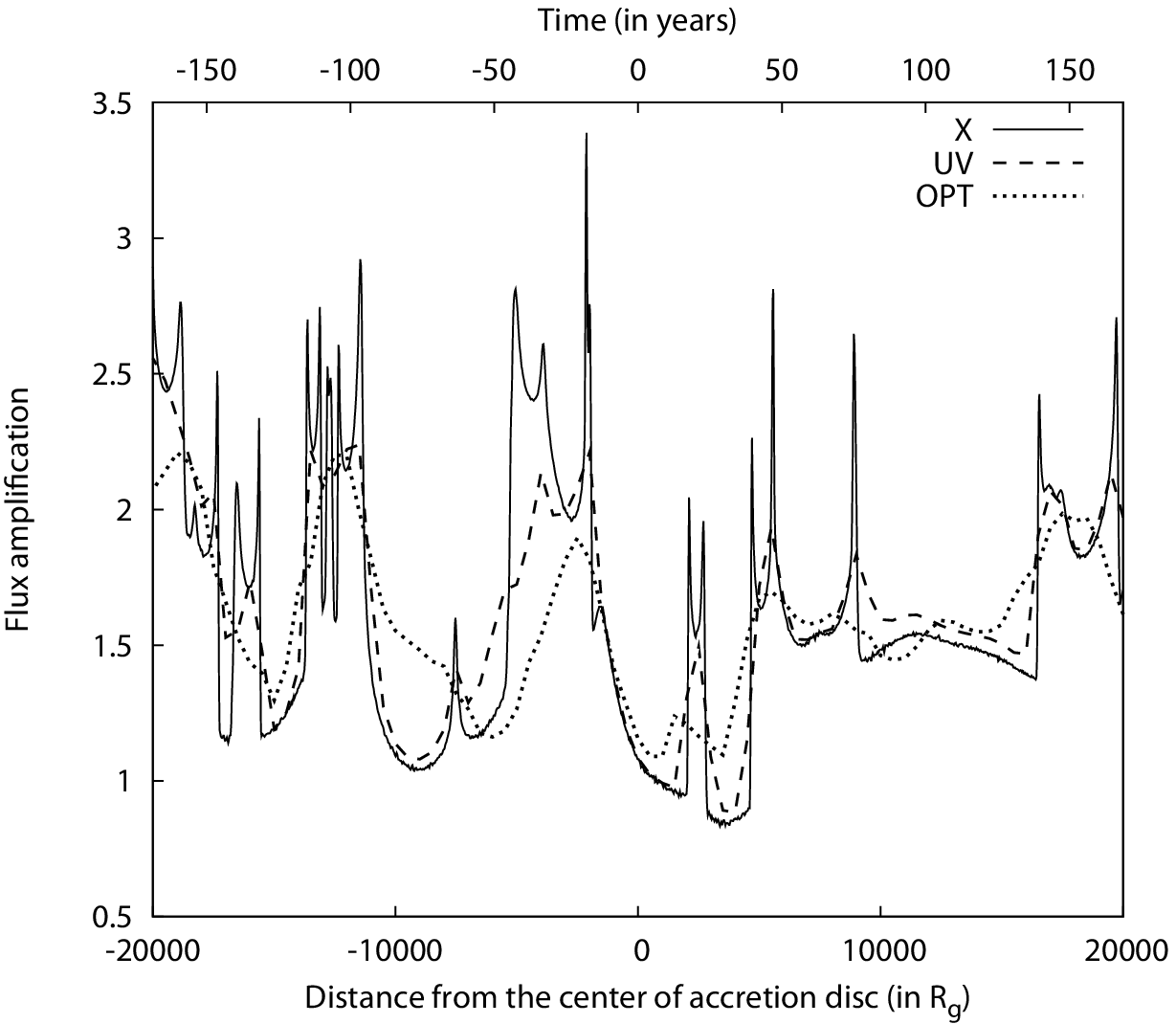}
\caption{The same as in Fig. 5 but for the magnification map of a
"typical" lens system  (Fig. 1 right).}
\end{figure*}

We also estimated time scales for seven lensed QSOs which have been
observed in the X-ray band \citep{Dai04}. For each spectral band two
estimates are made: caustic time ($t_{\rm caustic}$) - obtained from
equation (14) and caustic rise time ($t_{HME}$) - obtained from
caustic crossing simulations (see Figs. 2 -- 4). In the second case,
we measured the time from the beginning of the microlensing event
until it reaches its maximum (i.e. in the direction from the right
to the left in Figs. 2 -- 4). The duration of the magnification
event beyond the maximum of the amplification could not be
accurately determined because of the asymptotic decrease of the
magnification curve beyond the peak. Estimated time scales for
different spectral bands are given in Table 1. As one can see from
Table 1, the microlensing time scales are significantly smaller for
the X-ray than for the UV/optical bands.

Unamplified and amplified brightness profiles of the X-ray emitting
region, corresponding to the highest peak in Fig. 5 (top right) are
presented in Fig. 7. As one can see in Fig. 7, the assumed
brightness profile of the source is very complex due to applied ray
tracing method, which allows us to obtain an image of the entire
disc, not only its one dimensional profile. Therefore, we could not
use a simple source profile for the calculation of microlensing time
scales (as it was done by \citet{Witt95}). Instead, we estimated the
frequency of high amplification events (HMEs), i.e. the number of
such events per unit length, directly from the light curves
presented in Figs 5 and 6. For models with non-zero shear, this
frequency depends on the direction of motion and for both calculated
maps (Q2237+0305A and "typical" lens) we counted the number of high
magnification events along the following paths (see Fig. 1): i)
horizontal ($y=0$) in the direction from $-x$ to $x$, ii) diagonal
($y = -x$) in direction from $-x$ to $x$ and iii) vertical ($x=0$)
in direction from $y$ to $-y$. For each map the lengths of the
horizontal and vertical paths in the source plane are the same
(13.636 $R_E$ for Q2237+0305A and 12.875 $R_E$ for "typical" lens),
as well as are the corresponding crossing times (66.22 years for
Q2237+0305A and 337.32 years for "typical" lens). For Q2237+0305A
the length of the diagonal path in source plane is 19.284 $R_E$ and
the crossing time is 93.62 years, while the corresponding length in
the "typical" case is 18.208 $R_E$ and the crossing time is 477.04
years.

\begin{figure*}
\centering
\includegraphics[width=0.495\textwidth]{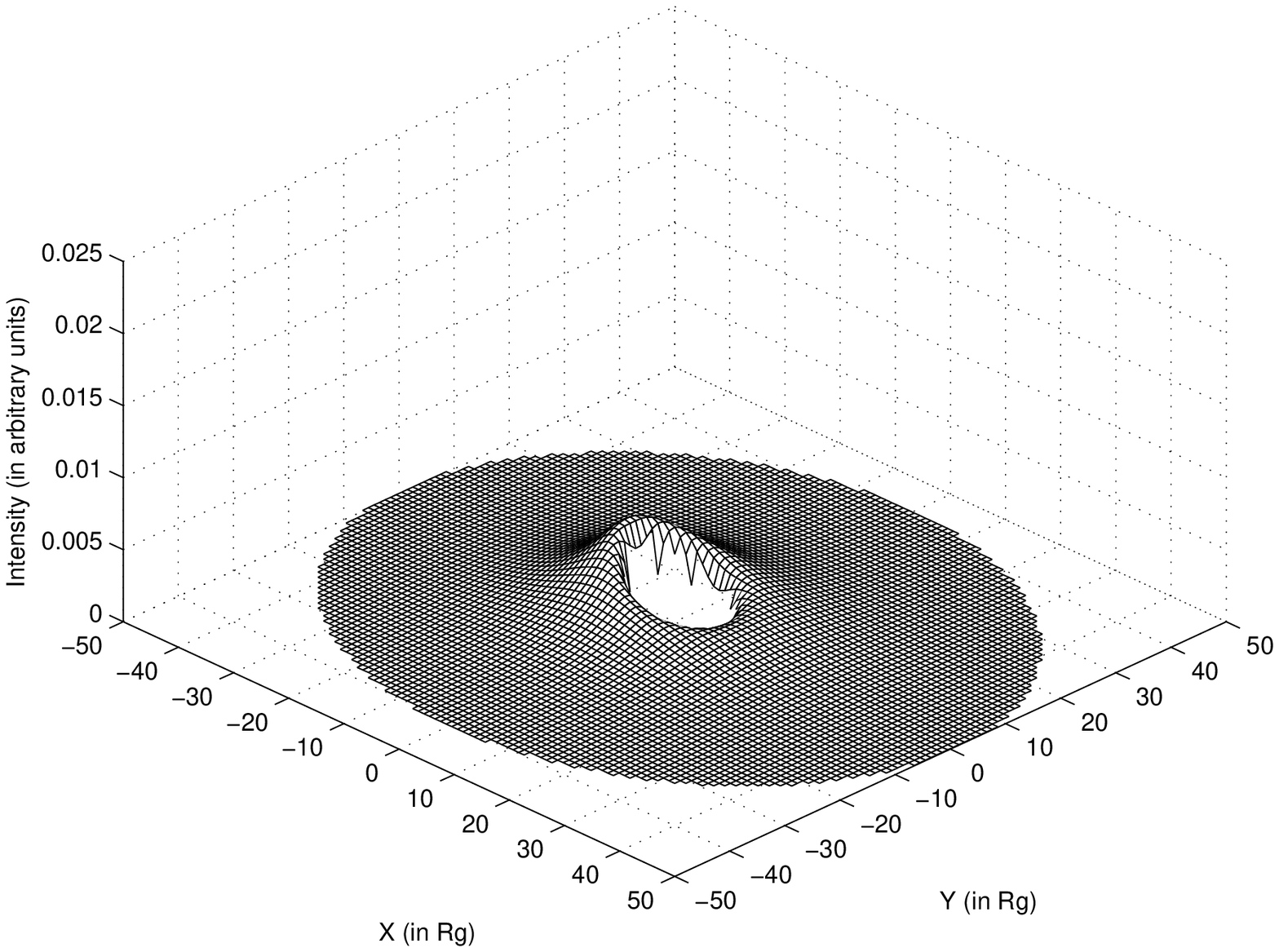}
\includegraphics[width=0.495\textwidth]{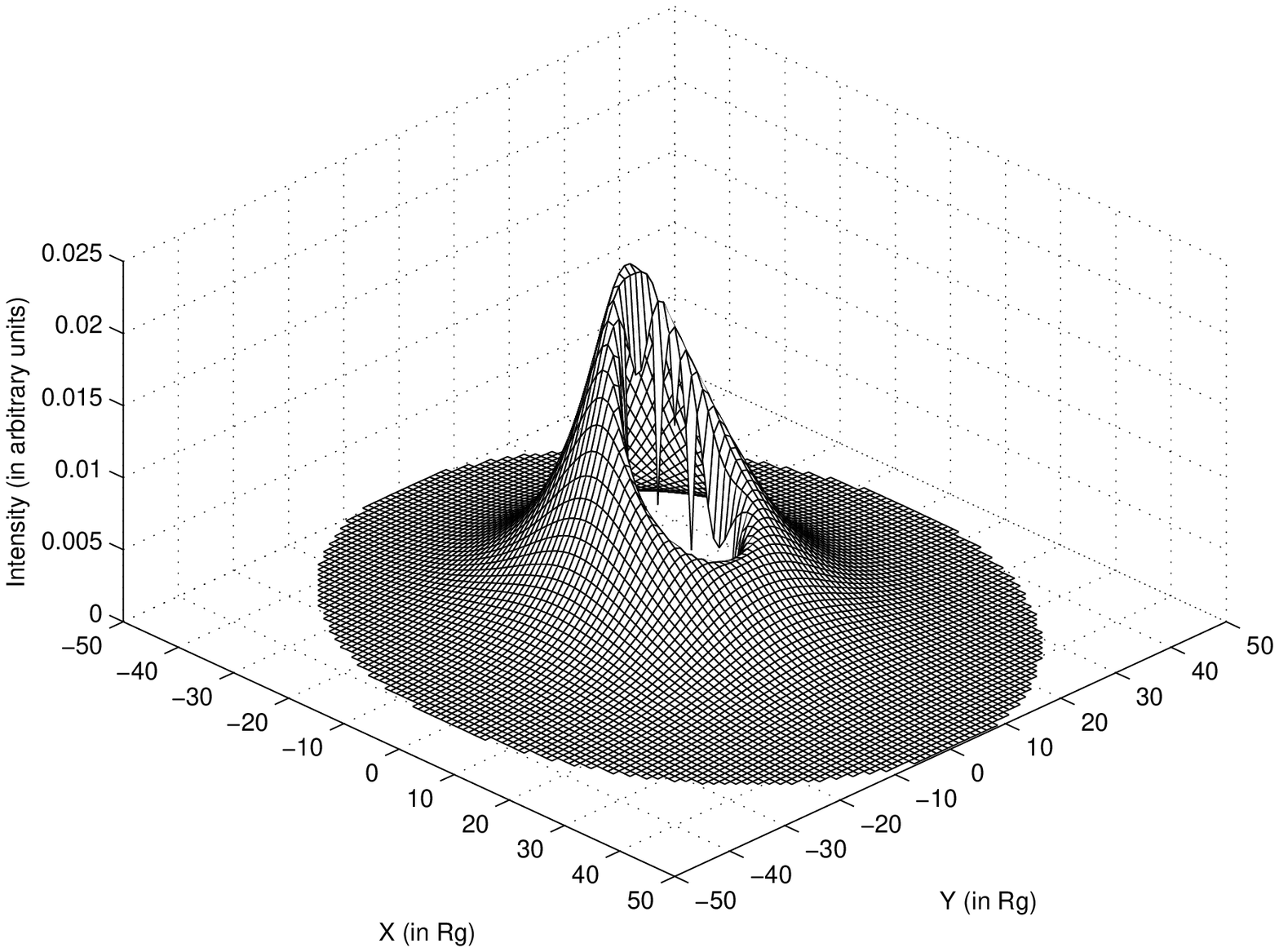}
\caption{Unamplified (left) and amplified (right) brightness profile
of the X-ray emitting region, corresponding to the highest peak in
Fig. 5 (top right). The profiles are obtained using the ray tracing
method (see, e.g., Popovi\'c et al. 2003a,b and references
therein).}
\end{figure*}

HMEs are asymmetric peaks in the light curves which depend not only
on microlens parameters but also on the sizes of emitting regions in
the following sense: the larger emitting regions are expected to
produce smoother light curves and more symmetric peaks
\citep{Witt95}. Consequently, it can be expected that the majority
of HMEs should be detected in X-ray light curves, less of them in UV
and the smallest number in optical light curves. Therefore, we
isolated only clearly asymmetric peaks in all light curves and
measured their rise times $t_{HME}$ as the intervals between the
beginning and the maximum of the corresponding microlensing events.
In the case of Q2237+0305A we found the following number of HMEs: i)
horizontal path: 13 in X-ray, 8 in UV and 7 in optical band, ii)
diagonal path: 10 in X-ray, 7 in UV and 5 in optical band and iii)
vertical path: 22 in X-ray, 12 in UV and 6 in optical band. In case
of "typical" lens these numbers are: i) horizontal path: 21 in
X-ray, 10 in UV and 8 in optical band, ii) diagonal path: 30 in
X-ray, 9 in UV and 7 in optical band and iii) vertical path: 18 in
X-ray, 7 in UV and 4 in optical band.

\begin{figure*}
\centering
\includegraphics[width=0.495\textwidth]{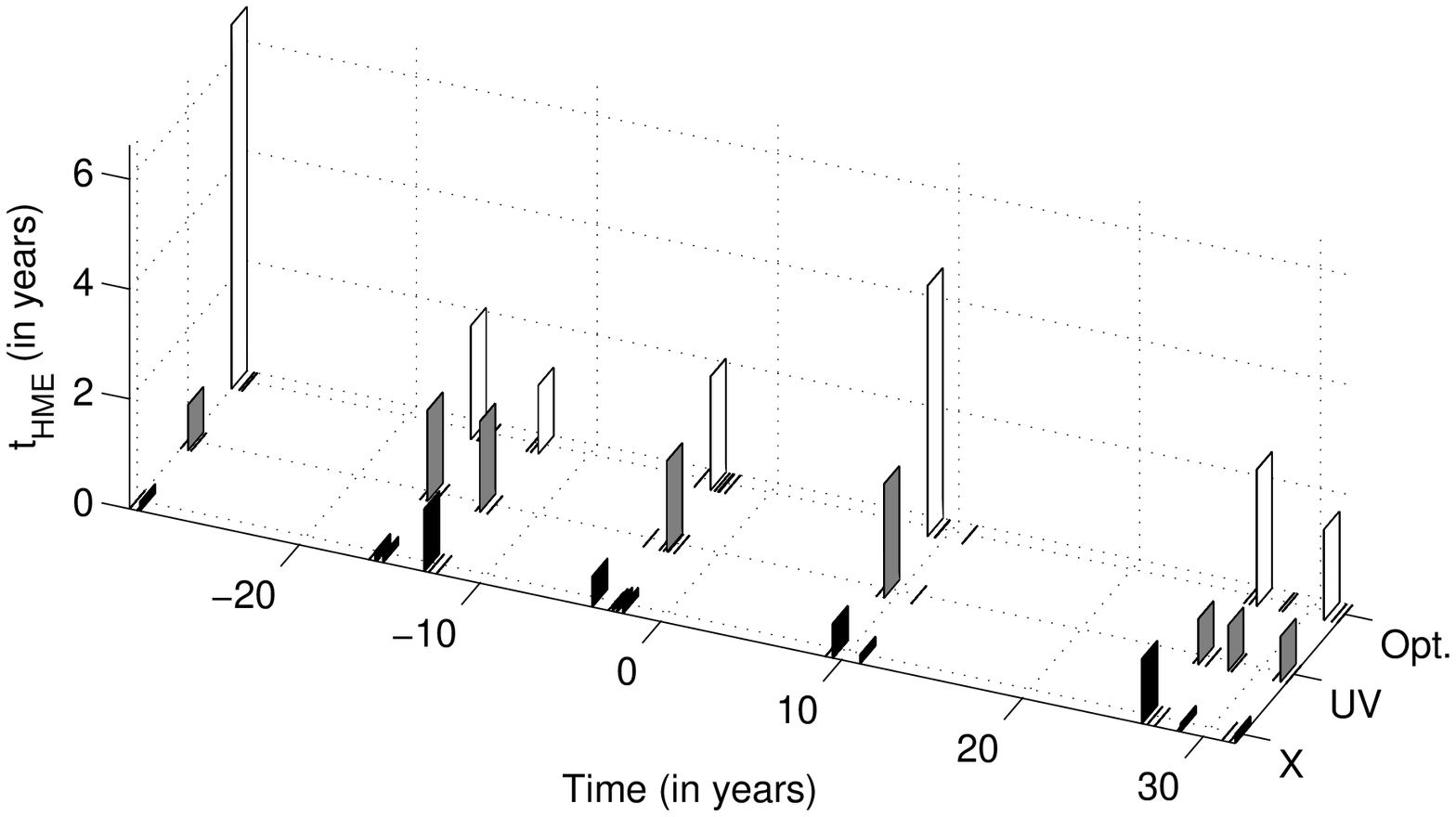}\hfill
\includegraphics[width=0.495\textwidth]{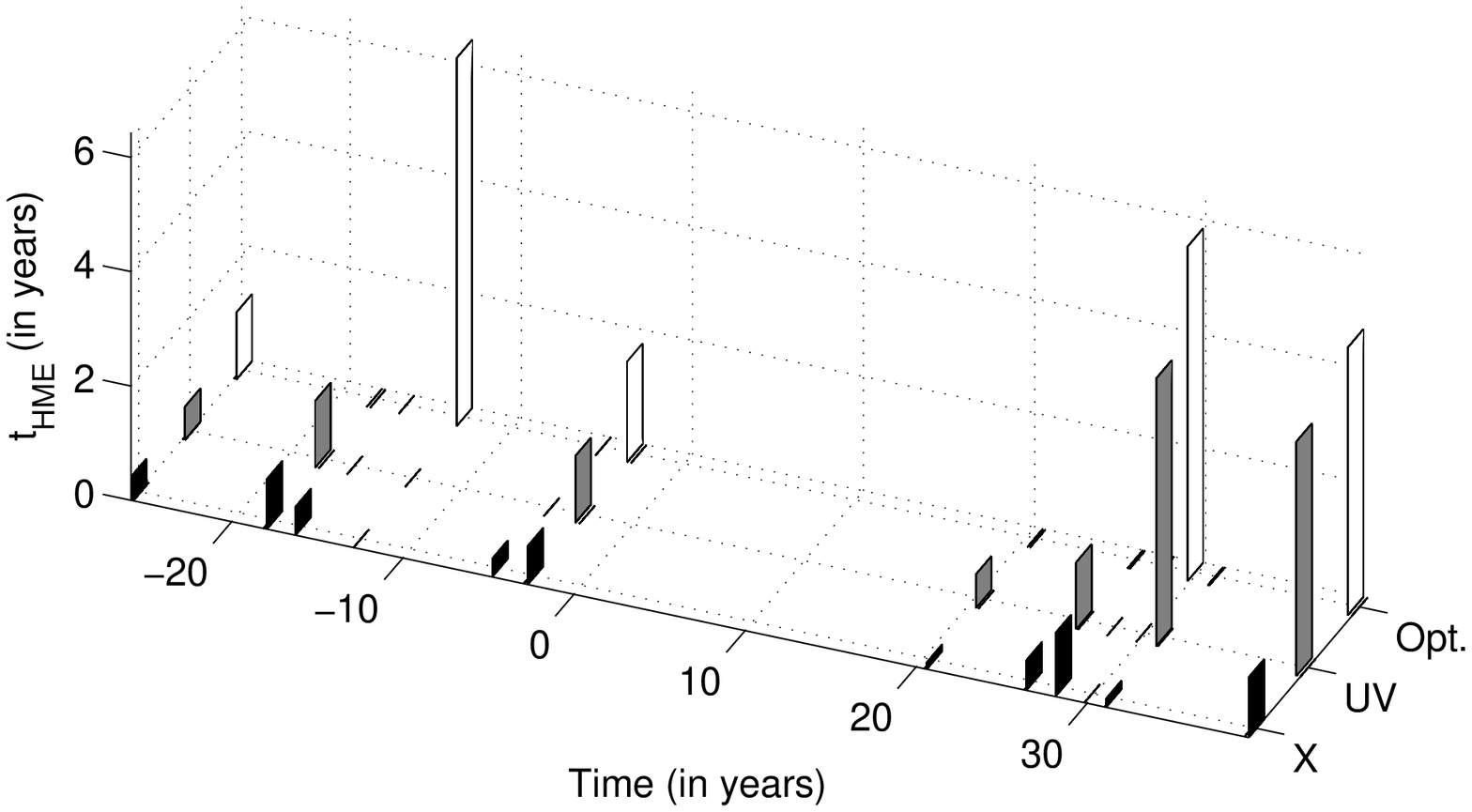} \\
\vspace*{0.3cm}
\includegraphics[width=0.495\textwidth]{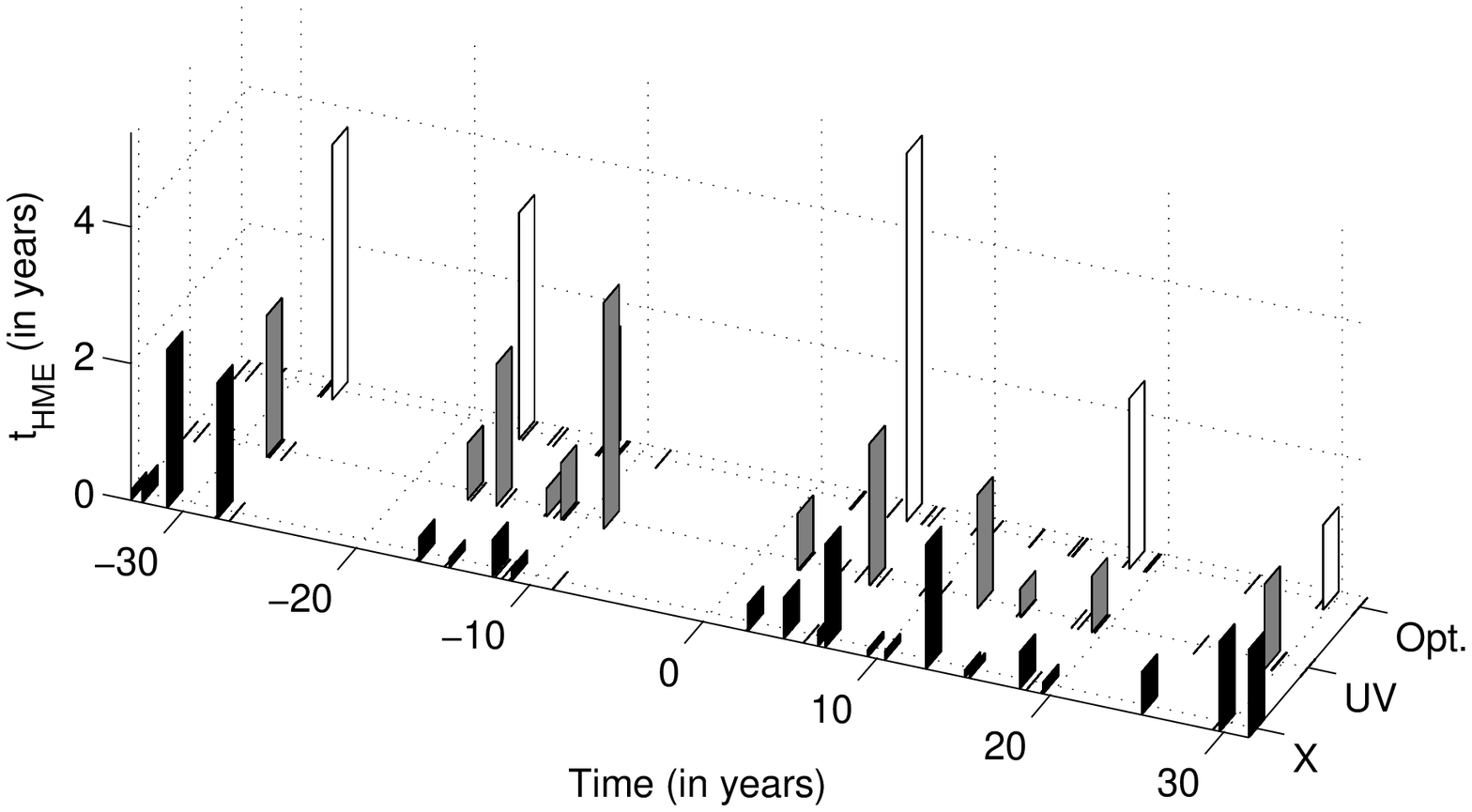}
\caption{Rise times ($t_{HME}$) of high magnification
events for all three spectral bands in the light curves of
Q2237+0305A (Fig. 5). Top left panel corresponds to the
horizontal, top right to the diagonal and bottom to the
vertical path of accretion disc.}
\end{figure*}

\begin{figure*}
\centering
\includegraphics[width=0.495\textwidth]{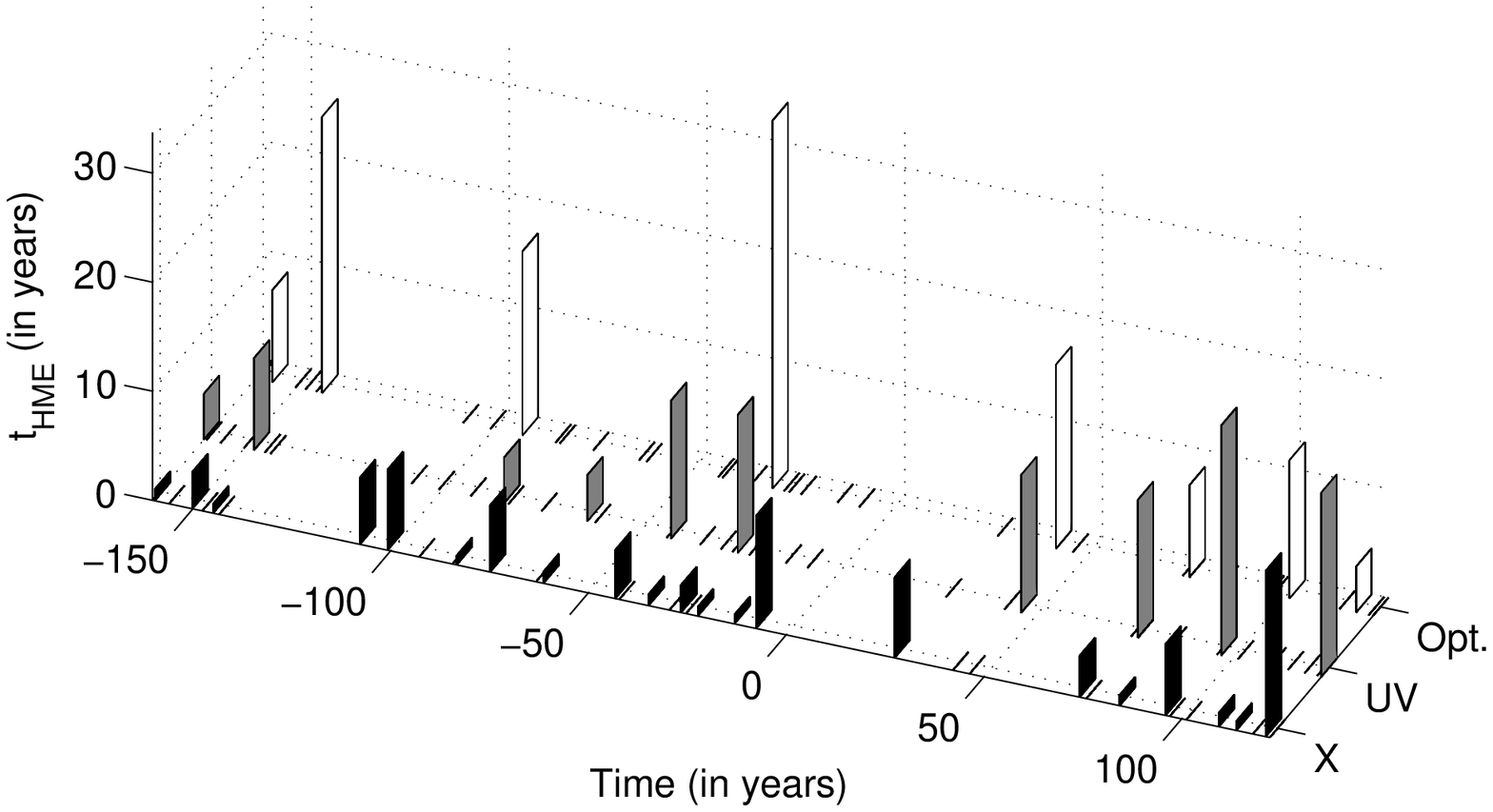}\hfill
\includegraphics[width=0.495\textwidth]{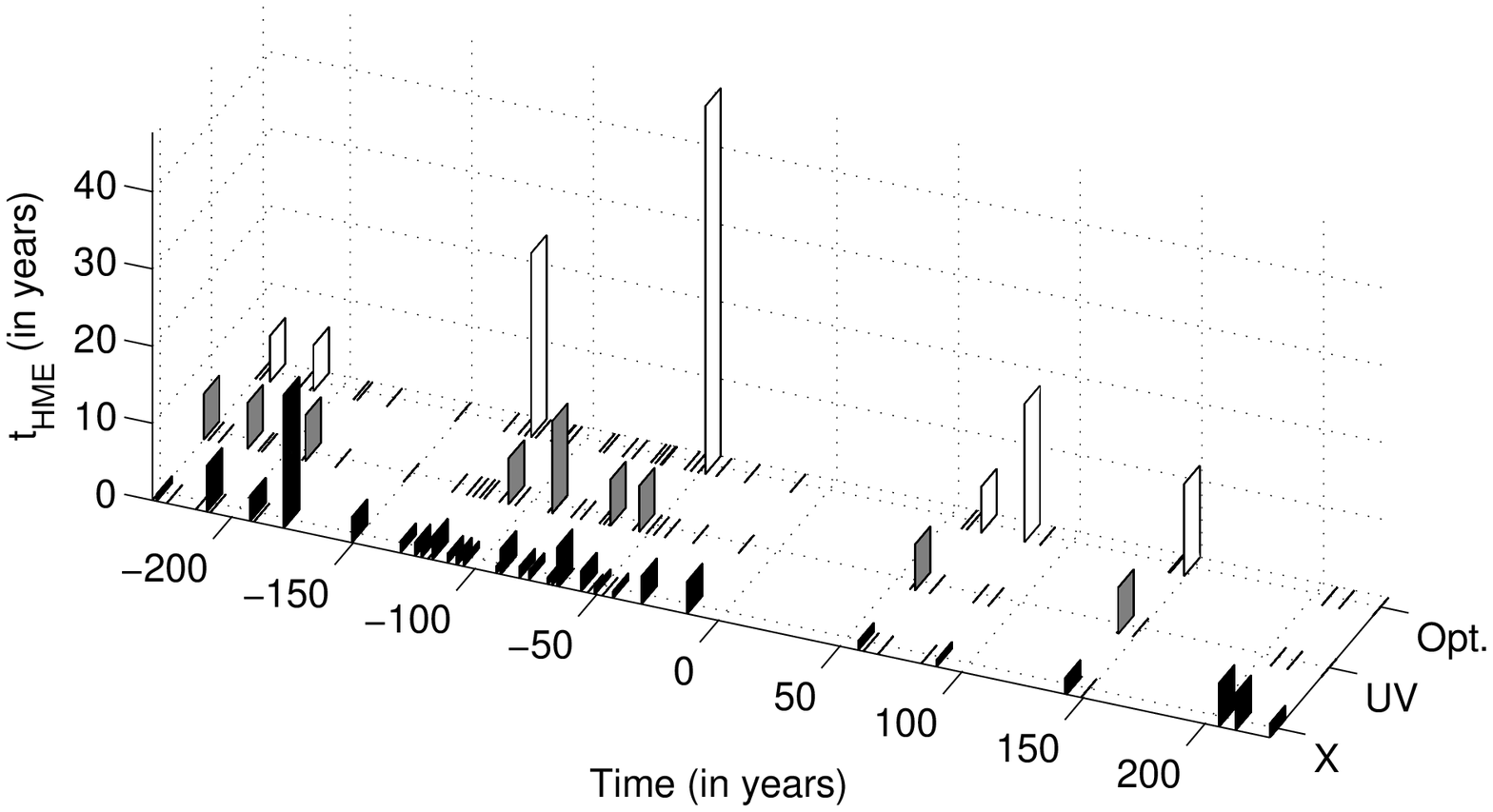} \\
\vspace*{0.3cm}
\includegraphics[width=0.495\textwidth]{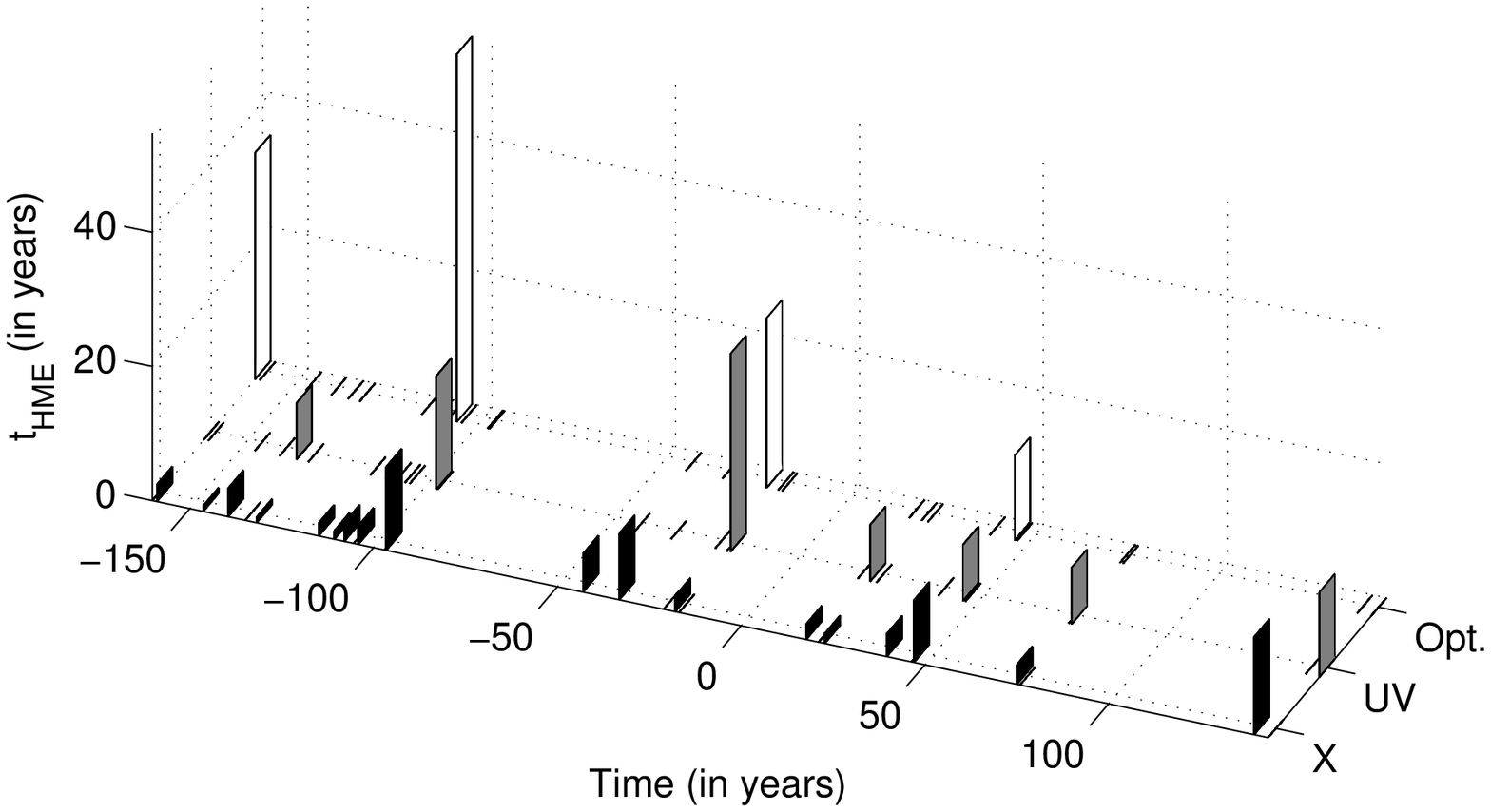}
\caption{The same as in Fig 8. but for the light
curves of "typical" lens system (Fig. 6)}
\end{figure*}

The average number of caustic crossings per unit length ($<N_{\rm
caustic}>_0$) and per year ($<N_{\rm caustic}>_y$) are given in
Table 2. This table also contains the average rise times
($<t_{HME}>$) in all three spectral bands, derived from the rise
times ($t_{HME}$) of individual HMEs which are presented in Figs. 8
and 9 in the form of histograms. These results, as expected, show
that the rise times are the shortest and the frequency of caustic
crossings is the highest in the X-ray spectral band in comparison to
the other two spectral bands. One can also see from Tables 1 and 2
that in the case of Q2237+0305A, the average rise times of HMEs for
all three spectral bands (obtained from microlens magnification
pattern simulations) are longer than both, caustic rise times
(obtained from caustic simulations) and caustic times (calculated
from equation (14)).

Microlensing can result in flux anomalies in the sense that
different image flux ratios are observed in different spectral bands
\citep{Pc04,Pop06b}. As shown in Figs 2--6. the amplification in the
X-ray band is larger and lasts shorter than it does in the UV and
optical bands. Consequently, monitoring of lensed QSOs in the X-ray
and UV/optical bands can clarify whether the flux anomaly is
produced by CDM clouds, massive black holes or globular clusters
(millilensing) or stars in foreground galaxy (microlensing).

%\clearpage

\section{Conclusion}

In this paper we calculated microlensing time scales of different
emitting regions. Using a model of an accretion disc (in the center
of lensed QSOs) that emits in the X-ray and UV/optical spectral
bands, we calculated the variations in the continuum flux caused by
a straight-fold caustic crossing an accretion disc. We also
simulated crossings of accretion discs over microlensing
magnification patterns for the case of image A of Q2237+0305 and for
a "typical" lens system. From these simulations we concluded the
following:

(i) one can expect that the X-ray radiation is more amplified than
UV/optical radiation due to microlensing which can induce the so
called 'flux anomaly' of lensed QSOs.

(ii) the typical microlensing time scales for the X-ray band
are on order of several months, while for the UV/optical they
are on order of several years (although the time scales
obtained from microlensing magnification pattern simulations
are longer in comparison to those obtained from caustic
simulations).

(iii) monitoring of the X-ray emission of lensed QSOs can
reveal the nature of 'flux anomaly' observed in some lensed
QSOs.

All results obtained in this work indicate that monitoring the X-ray
emission of lensed QSOs is useful not only to discuss the nature of
the 'flux anomaly', but also can be used for constraining the size
of the emitting region.

\section*{Acknowledgments}

This work is a part of the project (146002) "Astrophysical
Spectroscopy of Extragalactic Objects" supported by the Ministry of
Science of Serbia. The authors would like to thank the anonymous
referee for very useful comments.

%\clearpage

\label{lastpage}

\end{document}